\documentclass[aps,prb,twocolumn,amsmath,amssymb,superscriptaddress,floatfix]{revtex4-2}

\usepackage{graphicx} 
\usepackage{mathptmx}
\usepackage{amsmath}
\usepackage[]{graphicx}      
\usepackage{bm}             	
\usepackage{times}			
\usepackage{tabularx}
\usepackage{color}
\usepackage{dcolumn}		
\usepackage{amsmath}
\usepackage{amssymb}
\usepackage{amsfonts}
\usepackage{calc}
\usepackage{times}
\usepackage{tabularx}
\usepackage{xcolor,colortbl}
\usepackage{rotating}
\usepackage[colorlinks,citecolor=blue,linkcolor=blue]{hyperref}
\usepackage{amsmath}
\usepackage{hyperref} 

\begin{document}

\title{
Symmetry of the coupling between surface acoustic waves and spin waves in synthetic antiferromagnets}

\author{R. Lopes Seeger}
\affiliation{Universit\'e Paris-Saclay, CNRS, Centre de Nanosciences et de Nanotechnologies, 91120, Palaiseau, France}
\affiliation{SPEC, CEA, CNRS, Université Paris-Saclay, 91191 Gif-sur-Yvette, France}
\author{L. La Spina}
\affiliation{Université de Franche-Comté, CNRS, Institut FEMTO-ST, 26 rue de l’Epitaphe, 25000 Besançon, France}
\author{V. Laude}
\affiliation{Université de Franche-Comté, CNRS, Institut FEMTO-ST, 26 rue de l’Epitaphe, 25000 Besançon, France}
\author{F. Millo}
\affiliation{Universit\'e Paris-Saclay, CNRS, Centre de Nanosciences et de Nanotechnologies, 91120, Palaiseau, France}
\author{A. Bartasyte}
\affiliation{Université de Franche-Comté, CNRS, Institut FEMTO-ST, 26 rue de l’Epitaphe, 25000 Besançon, France}
\author{S. Margueron}
\affiliation{Université de Franche-Comté, CNRS, Institut FEMTO-ST, 26 rue de l’Epitaphe, 25000 Besançon, France}
\author{A. Solignac}
\author{G. de Loubens}
\affiliation{SPEC, CEA, CNRS, Université Paris-Saclay, 91191 Gif-sur-Yvette, France}
\author{L. Thevenard}
\author{C. Gourdon}
\affiliation{Institut des Nanosciences de Paris, Sorbonne Université, CNRS, UMR 7588, 4 place Jussieu, F-75005 Paris, France}
\author{C. Chappert}
\author{T. Devolder}
\affiliation{Universit\'e Paris-Saclay, CNRS, Centre de Nanosciences et de Nanotechnologies, 91120, Palaiseau, France}

\date{\today}

\begin{abstract}
Synthetic antiferromagnets host spin waves that are highly tunable. It is 
of practical interest to analyze the symmetry of their coupling to surface acoustic waves 
with the design of hybrid devices in view. 
For this we calculate the layer-resolved susceptibility tensor of a synthetic antiferromagnet, the effective magneto-elastic and magneto-rotation fields associated to a travelling elastic wave, and the power irreversibly transferred by the elastic wave to the magnetic layers. We consider Rayleigh-type surface acoustic waves: (a) that travel in an elastically isotropic, non-piezoelectric substrate, or (b) that propagate along the X direction at the surface of a Z-cut LiNbO$_3$ substrate, or (c) that are guided in a thin Z-cut LiNbO$_3$ film grown on a sapphire substrate. 
In particular, we show that the complementary angular dependencies of the acoustic and optical spin wave modes in synthetic antiferromagnets makes it possible to excite spin waves for any relative orientation of magnetization and acoustic wavevector.
In addition, we discuss the symmetries of the driving fields and of the energy transferred to the magnetic degree of freedom. We evidence new interaction channels coupling the magnetization eigenmodes when elastic anisotropy and piezoelectricity of the substrate are considered.
\end{abstract}

\maketitle

\section{Introduction}


The coupling between surface acoustic waves (SAWs) and spin waves (SWs) has drawn increasing attention in recent years due to its potential for the realization of integrated microwave devices benefiting from the best of both magnonics and microwave acoustics \cite{Weiler2011,Thevenard2014,Sasaki2017,Puebla2020,Hernandez2020,Xu2020}. Particularly, the coupling of SWs in magnetic thin films with SAWs allows excitation, manipulation and detection of various magneto-elastic waves. SAWs are excited on piezoelectric substrates, typically bulk single crystals of LiNbO$_3$.  The SAW-SW coupling is mediated by strains \cite{Weiler2011} and by the rotational motion of the lattice \cite{Maekawa1976} within the magnetic material.
However the SAW-SW coupling is generally effective only for certain relative orientations between magnetization and SAW wavevectors \cite{Dreher2012}. The coupling often vanishes in the high symmetry directions, except for specific substrates supporting Rayleigh and shear horizontal SAW modes simultaneously \cite{Kuss2021}.
Finding ways to couple SWs and SAWs for arbitrary orientations would enable to design hybrid devices that fully exploit the anisotropic and non-reciprocal properties of SWs as well as the high energy-efficiency of SAWs excitation and manipulation. 
In this paper, we considered an alternative magnetic material platform: the synthetic antiferromagnets (SAFs). For the analysis, we developed an extensive modeling of SAWs propagation and its interaction with SAF SWs.

The SAW-SW coupling mechanism is intrinsically non-reciprocal, because of an helicity mismatch \cite{Rasmussen2021,Hernandez2020,Xu2020,Kuess2020, devolder_propagating-spin-wave_2023}. In addition, non-reciprocal SAW transmission can also arise from an asymmetric SW dispersion relation \cite{Kuss2023,kus_giant_2023,verba2021phase,bas2022nonreciprocity}. Hybrid SW/SAW systems made from SAF may thus find applications, such that the SAW to SW coupling deserves to be studied. 
SAFs have two families of precession eigenmodes \cite{Zhang1994}: In-phase and out-of-phase precessions of the magnetization in each layer are associated with acoustic and optical SW modes, respectively. As we will show, the SAW to SW coupling displays different angular dependencies for each of the SAF eigenmodes.

To ease the modelling of SAW propagation, the substrate is usually assumed to be uncapped, elastically isotropic and its piezoelectric character is ignored. It, therefore, hosts an (idealized) Rayleigh SAW \cite{Weiler2011,Xu2020,Gowtham2015,Kuess2020,Sasaki2017,Tateno2020,Labanowski2016}. 
One often postulates that the magnetic layer has the same elastic properties as that of the substrate or, that the back action of the magnetic films on the substrate elastodynamics can be neglected.  The finite thickness of the magnetic system is then only translated as a "filling factor" in the SAW-SW coupling, save for rare occasions implementing the more complex "thin film on substrate approach" \cite{Rovillain2022,Gowtham2016}. In this article, we reconsider this first assumption, and consider the  impact of the substrate's elastic anisotropy and piezoelectricity onto the coupling of SWs to SAWs.

This paper presents a comprehensive symmetry analysis of the coupling between Rayleigh-type elastic waves traveling in various substrates with the SWs of a capping CoFeB/Ru/CoFeB SAF film. In particular we compare the case of a conventional Rayleigh wave in a semi-infinite elastically isotropic substrate ("iso" model), to the ones of a wave travelling in a real material (Z-cut LiNbO$_3$: "Z-cut" model) and of a wave guided in a Z-cut LiNbO$_3$ film grown on an Al$_2$O$_3$ C-sapphire substrate ("guided" model), see Fig.~\ref{xyzXYZ}(a). We clarify the respective roles of the magneto-elastic tickle fields, the magneto-rotation "rolling" fields and the two SW resonance modes of a SAF when set in the scissors state by an external magnetic field. Finally, we discuss interaction channels allowing SAW-SW coupling for all applied field orientations. 

The paper is organized as follows. We first assess the magnetics of SAF and its SW modes (section \ref{MagneticsOfaSAF}). We then compare the strain tensors and the rotation tensors of Rayleigh surface acoustic waves in the "iso", "Z-cut" and "guided" models in section \ref{SAWsection}.
The power transmitted to the spin waves excited by the elastic waves is studied in section \ref{sectionCoupling}. The amplitude of the different coupling channels are compared. Section \ref{SAWSW-iso} focuses then on a model system, the SAW to SW coupling in the fully isotropic case. The differences with more realistic materials represented by the other substrate models are finally discussed in section \ref{SAWSW-otherSubstrates}. 

\section{Geometry } 
We consider the geometry described in Fig.~\ref{xyzXYZ}. It consists in a synthetic antiferromagnetic film grown on substrate ending at $z=0$ that hosts surface acoustic waves within its depth ($z<0$). Two coordinate systems are defined. 
The coordinate system $\{XYZ\}$ is defined with respect to the wavevector $k_\textrm{SAW} \parallel X$ of the surface acoustic wave and the out-of-plane direction $Z$. This coordinate system will be used to describe the elastic properties and the surface acoustic waves. Three different substrates will be used:  a semi-infinite elastically isotropic substrate ("iso" model), a semi-infinite Z-cut LiNbO$_3$ single crystal ("Z-cut" model) and, a 150-nm thick Z-cut LiNbO$_3$ film grown on a semi-infinite C-sapphire substrate ("guided" model). It is noteworthy that $Y_{LNO} \perp (0\bar{1}10)_\textrm{sapphire}$. For the two anisotropic substrates, the $X$ direction matches with the $X$ crystalline direction of LiNbO$_3$. 
The full system is invariant in the direction transverse to the SAW wavevector (i.e. $\frac{\partial}{\partial Y} =0$).

The other coordinate system, i.e. $\{xyz\}$ is defined with respect to the growth direction $z=Z$ and a uniform static applied field $H_x$ oriented in the film plane at an angle $\varphi$ away from the SAW direction. This coordinate system will be convenient when describing the magnetics of the synthetic antiferromagnet. We shall consider that the SAF is thin enough and elastic enough to replicate perfectly to the strain present at the surface of the (much thicker) substrate, and not to have a back action on the substrate. 
To pass from one coordinate system to the other, it is useful to define the rotation matrix $\bar {\bar R}$ :
 \begin{equation}
\bar {\bar R}= 
    \left(
\begin{array}{ccc}
\cos \varphi& \sin \varphi& 0 \\ 
-\sin \varphi& \cos \varphi & 0 \\ 
0 & 0 & 1\\
\end{array}
    \right)
\end{equation}

\section{Modeling of the Synthetic Antiferromagnet under strain} \label{MagneticsOfaSAF}

\subsection{Magnetic energies} 
We model the SAF as two layers $\ell=1, 2$ of thickness $t_{\ell}$ and magnetizations $\left(m_{x\ell}, m_{y\ell}, m_{z\ell}\right)M_s $, the magnetizations being uniform across the thickness of each layer. The in-plane applied field $H_x$ leads to a Zeeman energy per unit surface that is written as $E_\textrm{Zeeman}=-  \mu_0  M_s H_{x} \sum_{\ell=1,2} m_{x\ell} \,t_\ell$. The areal shape anisotropy of the system \footnote{The intralayer exchange energy and the non-uniform demagnetizating fields arising at finite SW wavectors are disregarded} is $E_\textrm{shape}=+ \frac{1}{2} \mu_0 M_s^2 \sum_{\ell=1,2} m_{z\ell}^2 \,t_\ell $, and the areal energy due to the interlayer coupling can be written as $E_\textrm{J}=- J \left(m_{x1} m_{x2}+m_{y1} m_{y2}+m_{z1} m_{z2} \right)$. In the presence of mechanical deformations, there is also a magneto-elastic contribution written as \cite{Dreher2012} \footnote{Note that we neglect the part of the magneto-crystalline anisotropy that does not originate from magnetoelasticity.}
\begin{equation} \begin{split} E_\textrm{mel}=B_1  \sum_{\ell=1,2} t_\ell \left(m_{x\ell}^2 {\epsilon _{xx}}+m_{y\ell}^2 {\epsilon  _{yy}}+m_{z\ell}^2 {\epsilon _{zz}}\right) \\ 
+2 B_2 \sum_{\ell=1,2} t_\ell \left(m_{x\ell} m_{y\ell} {\epsilon  _{xy}}+m_{x\ell} m_{z\ell} {\epsilon _{xz}}+m_{y\ell} m_{z\ell} {\epsilon  _{yz}}\right) \\
\end{split} \end{equation} and a magneto-rotation contribution \cite{Xu2020}: \begin{equation}E_\textrm{roll}=+\mu_0 M_{s}^2 \sum_{\ell=1,2} t_\ell m_{z\ell}  (m_{x\ell} {\omega _{xz}}+m_{y\ell}
   {\omega _{yz}}),\end{equation}where $ \bar {\bar \epsilon}(X, Z, t) $ is the strain tensor and $\bar {\bar \omega}(X, Z, t)$ is the lattice rotation tensor within the SAF. $B_1$ and $B_2$ are the magneto-elastic coefficients. It is noteworthy that the rotational lattice motion of the SAWs causes reorientation of the surface normal direction, which couples to the magnetization via magnetic anisotropy fields. Therefore, the magneto-rotation contribution is also active in materials with vanishing magnetostriction.
Note that since we assume easy plane with isotropic magnetic properties, the directions within the $xy=XY$ planes are magnetically equivalent. The rotation $\omega_{xy}$ (i.e. rotation about the $z$ axis) thus does not contribute to the magnetic energy. The lattice deformations lead to the effective "tickle" $h_{\textrm{tick}}^{rf}$ and "rolling" $h_\textrm{{roll}}^{rf}$ driving fields, as we will discuss further below. 

In the absence of data in the literature, we shall follow the usual speculation and assume an isotropic magnetic system which leads to $B_1=B_2$.
Typical values for Co$_{40}$Fe$_{40}$B$_{20}$ include a magnetization \cite{mouhoub_exchange_2023} $M_s=1.35$ MA/m, an interlayer exchange coupling \cite{seeger_inducing_2023} $J= -1.1~\textrm{mJ/m}^2$ for a $0.7$ nm thick Ru spacer, a first magneto-elastic constant \cite {Masciocchi_strain_2021} $B_1=-7.6~\textrm{MJ/m}^3$, which for layer thicknesses $t_1=t_2= 13~\textrm{nm}$ 
lead to characteristics fields of $\mu_0 M_s=1.7$ T, 
$\mu_0 H_{\textrm{mel}}=\frac{2B_1} {M_s}= 11.2~\textrm{T}$ and interlayer exchange field $\mu_0 H_J=-\frac{2 J}{M_s t_{\ell}}= 125~\textrm{mT}$. 

\begin{figure} 
\hspace{-6mm}
\includegraphics[width=7.2cm]{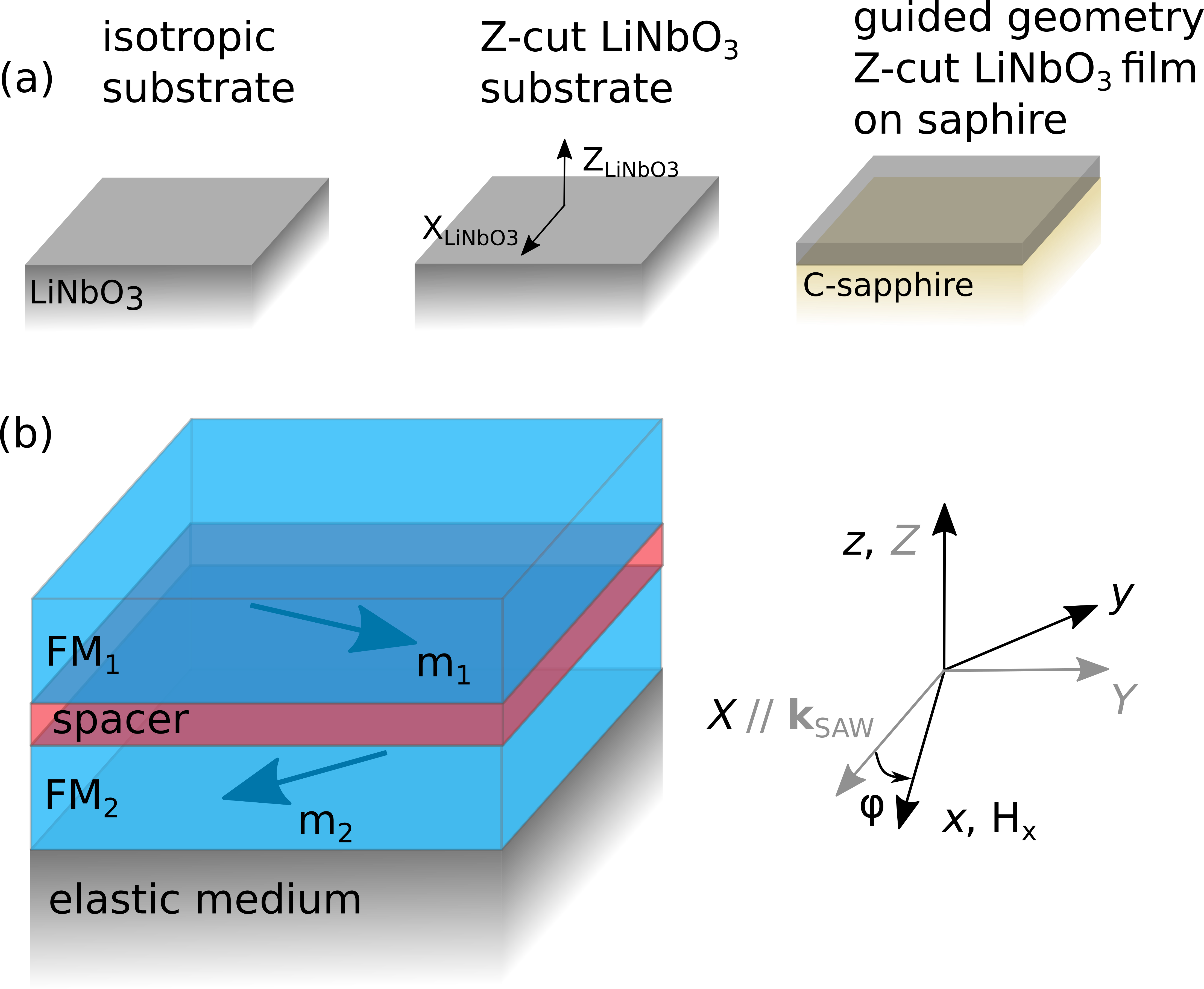}
\caption{Schematics of the investigated systems. (a) Studied elastic media. (b) The SAF is one of the two degenerate scissors state. The coordinate system $\{xyz\}$ is defined with respect to the growth direction $z=Z$ and the static applied field $H_x$ oriented in the film plane. The coordinate system $\{XYZ\}$ is defined with respect to the wavevector of the surface acoustic wave $k_\textrm{SAW} \parallel X$ and the out-of-plane direction $Z$.
}
\label{xyzXYZ}
\end{figure}

\subsection{Ground state of the synthetic antiferromagnet} \label{groundstate}

In order to allow for analytical calculations we shall simplify the micromagnetic problem. We consider that the ground state of the SAF is equal to that of a two-macrospin system. In other words, we neglect any gradient of the magnetization within the thickness of the magnetic layers \cite{mouhoub_exchange_2023}. This assumption will be lifted temporarily in Fig.~\ref{saf}(c) and in the related discussion (section \ref{bemol}).   

We consider that the applied field follows $0 \leq H_x < H_J$, and that the rf elastic deformations have vanishing time-averages 
such that at rest the SAF is in one of the two degenerate scissors states. We choose the one for which the layers are magnetized as: \begin{equation}\left( \begin{array}{c}  m_{x\ell} \\ m_{y\ell} \\ m_{z\ell} \\ \end{array} \right) = 
\left( \begin{array}{c} \frac{H_x}{H_J} \\ \delta \sqrt{1- \frac {H_x^2}{H_J^2}} \\ 0 \\ \end{array} \right) 
 \label{GroundState} \end{equation}
where $\delta=1$ for the layer $\ell=1$ and $\delta=-1$ for layer $\ell=2$.

\subsection{Tickle fields and rolling fields} 
The description of magnetization dynamics requires to depict the effective fields acting on the magnetization in the frame of the applied field, i.e., $xyz$ [Fig.~\ref{xyzXYZ}(b)]. 
The SAW induced lattice deformations lead to the so-called magneto-elastic "tickle" fields \footnote{Note that for an in-plane magnetized system, $\epsilon_{zz}$ does not induce a torque at first order.}:
\begin{equation}
h_\textrm{tick}^{rf}= - \frac{2}{\mu_0 M_s}
\left(
\begin{array}{c}
  B_1 \frac{H_x}{H_J} \,\epsilon_{xx}+ B_2
 \delta \sqrt{1-\frac{H_x^2}{H_J^2}} \,\epsilon_{xy} \\
 B_2\frac{H_x}{H_J} \,
\epsilon_{xy}+ B_1
   \delta \sqrt{1-\frac{H_x^2}{H_J^2}} \, \epsilon_{yy} \\
  B_2\frac{H_x}{H_J} \,
\epsilon_{xz}+ B_2 
   \delta \sqrt{1-\frac{H_x^2}{H_J^2}}\, \epsilon_{yz} \\
\end{array}
\right)
    \label{TickleFields}
\end{equation}
as well as to the magneto-rotation "rolling" fields:
\begin{equation}
 h_\textrm{roll}^{rf}= M_{s}\left( \begin{array}{c}
0 \\  0 \\ \omega_{zx}\frac{H_x
    }{H_J} ~+ ~\omega_{zy} \delta \sqrt{1-\frac{H_x^2}{H_J^2}} \\
\end{array} \right) \label{RollingFields}
\end{equation}
Note that owing to the $\delta$ terms in Eq.~\ref{TickleFields} and \ref{RollingFields}, these effective fields are layer-dependent in a SAF. When a SAW is present, these effective fields are also space and time dependent and their values can be found by the substitution $ h^{rf} \rightarrow \mathcal{R}e ( h^{rf} e^{i(\omega t - k_\textrm{SAW} X)})$.

\subsection{Spin waves of the synthetic antiferromagnet} 
\subsubsection{Uniform resonances} 
To get the \textit{uniform} resonances of the SAF, we consider the total energy of the system and linearize the Landau-Lifschitz-Gilbert equation around the chosen ground state. We follow the standard procedure and write the dynamical matrix of the system, whose eigenvalues are the two uniform resonance frequencies of a SAF [Fig.~\ref{saf}(a,b)].  
For $H_x < H_J$, their real parts read \cite{devolder2022population}:
\begin{equation}
 \frac {\omega_\textrm{ac}}{\gamma_0}=   H_{x} \sqrt{\frac{M_s+H_{J}}{H_{J}}} \textrm{~and~} \frac{ \omega_\textrm{op}}{\gamma_0} =  \sqrt{\frac{M_s}{H_{J}}} \sqrt{H_{J}^2-H_{x}^2},
 \label{FMRfrequencies} \end{equation} where the labels recall the acoustic (in-phase) or optical (out-of-phase) nature of the spin wave mode.
 The imaginary parts of the eigenvalues lead to the mode linewidths (full widths at half maximum): \begin{equation}
 \frac {\Delta \omega_\textrm{ac}}{\alpha \gamma_0 } = M_s + \frac{H_J^2+H_x^2}{H_J}~ \textrm{and~}  \frac {\Delta \omega_\textrm{op}}{\alpha \gamma_0 } = M_s + \frac{H_J^2-H_x^2}{H_J}, \end{equation}
where $\gamma_0$ is the gyromagnetic ratio and $\alpha$ is the damping.

\subsubsection{Working assumption: spinwaves with vanishing group velocity} \label{bemol}
\begin{figure} 
\hspace{-6mm}
\includegraphics[width=7.2cm]{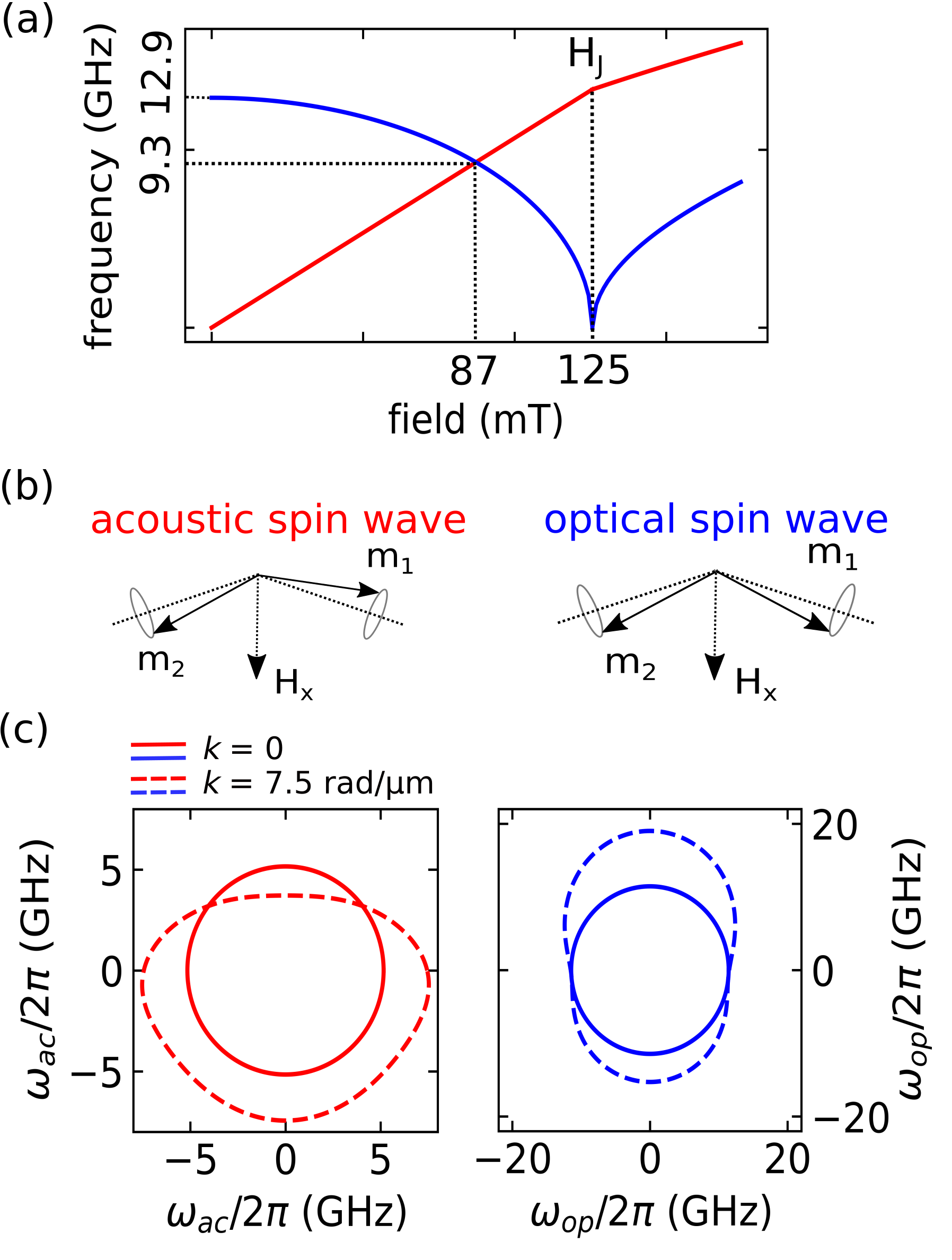}
\caption{(a) Acoustic and optical SW frequencies in a SAF from Eq.~ \ref{FMRfrequencies} for $H_x < H_J$ and extracted from Ref. \cite{devolder2022population} for $H_x > H_J$. (b) sketch of their respective magnetization precessional motions in the scissors state. (c) Angle-resolved dispersion relations $\omega_\textrm{mode}(\varphi)$ for $\mu_0 H_x= 50$ mT, $k = 0$ and 7.5 rad/µm within the full micromagnetics framework. 
}
\label{saf}
\end{figure}

We now discuss an important simplifying assumption. We consider that the spin waves have \textit{vanishing} group velocities, i.e. we use the definition:
\begin{equation} \omega_\textrm{ac, op}(k, \varphi) ~~\triangleq~~ \omega_\textrm{ac, op}(k=0, \varphi) \label{WorkingAssumption} \end{equation} 
The crude \cite{Ishibashi2020, Gallardo2019, millo_unidirectionality_2023-1} assumption of Eq.~\ref{WorkingAssumption} is done at the sake of pedagogy but it has an important consequence [Fig.~\ref{saf}(c)] that deserves a discussion. 
In reality the SW mode frequencies $\omega_{\textrm{mode}}$ for $k = 0$ and $k \neq 0$ and their angular dependencies with $\varphi$ differ. This is shown in Fig.~\ref{saf}(c), in which $\omega(k = 0, \varphi)$ from Eq.~\ref{FMRfrequencies} is compared with $\omega(k \neq 0, \varphi)$ when calculated in the full micromagnetic framework, as conducted in Ref. \cite{millo_unidirectionality_2023-1}. The non-circular character of $\omega(\varphi, k \neq 0)$ reflects the anisotropic character of the SWs of a SAF. The non-centro-symetric character of $\omega(\varphi, k \neq 0)$ recalls their  non-reciprocal nature \cite{Gallardo2019, Ishibashi2020, millo_unidirectionality_2023-1, devolder_propagating-spin-wave_2023}.

In a real experiment, there would be a defined SAW wavevector $k_\textrm{SAW}$ and a defined magnetic field orientation $\varphi$. Within our working assumption (Eq.~\ref{WorkingAssumption}), if the field amplitude $H_x$ is set to have a given SW mode be at resonance with the SAW, the field can be rotated while maintaining the resonant character. The angular dependence of the SAW-SW coupling then only reflects the angular dependence of the driving torque. 

In contrast with our working assumption, if the true dispersion of the spin wave was taken into account, rotating the field would automatically \textit{detune} the SW resonances away from the SAW resonance. As a result, the angular dependence of the SAW-SW coupling would reflect \textit{both} the angular dependence of the driving torque (so-called helicity mismatch \cite{devolder_propagating-spin-wave_2023}) \textit{and} the angular dependence of the SW frequencies (mismatch between the SAW resonance and the SW resonance).  

We aim to unravel the respective effects of angle-dependent resonance mismatch on the one hand, and, on the other hand, angle-dependent helicity mismatch. For that purpose our study will be conducted by assuming spin waves with vanishing group velocities (Eq.~\ref{WorkingAssumption}) to highlight only the SAW-SW coupling symmetry from the driving field standpoint.
Fully taking into account that $\omega_\textrm{ac, op}(k, \varphi) \neq \omega_\textrm{ac, op}(k\neq 0, \varphi)$ is an interesting subject that will be explored in future investigations.

\subsection{Susceptibility tensor of a SAF in the reference frame of the applied field} \label{SusceptibilitySection}
From the dynamical matrix, a cumbersome algebra leads to the susceptibility of each layer of the SAF for a zero wavevector.  
The SAF susceptibility can be separated into two independent contributions corresponding to the acoustic and optical eigenmodes, and it is expressed as:
\begin{equation}
\bar {\bar {\chi}}_{\ell} = \mathcal{L}^\textrm{ac} \left( \begin{array}{ccc}  0 &  \lambda \delta  & i \eta  \frac {\lambda} {\delta} \textcolor{black}{\frac{\omega}{\omega_\textrm{ac}}} \\  0 & 1 & i \eta \frac{\omega}{\omega_\textrm{ac}}  \\  0 & -i \eta \frac{\omega}{\omega_\textrm{ac}} & \eta ^2 \\ \end{array} \right)+ \mathcal{L}^\textrm{op} \left( \begin{array}{ccc}  1 & 0 & 0 \\  \frac{\delta}{\lambda } & 0 & 0 \\  -i \frac{\nu}{\delta} \textcolor{black}{\frac{\omega}{\omega_\textrm{op}}} & 0 & 0 \\ \end{array} \right) \label{ChiTotal}
\end{equation}
where the dimensionless terms $\eta$, $\lambda$ and $\nu$ describing the precession ellipticity are: $\eta = \sqrt{\frac{H_J}{H_J+M_s}}$,~$ \lambda=\frac{\sqrt{H_J^2- H_x^2}}{H_x}$ and $\nu= \sqrt{\frac{H_J}{M_s}}$. Note the non-hermitian character of the layer-resolved $\bar {\bar {\chi}}$. The Hermitian character (reflecting the gyrotropy) is recovered when summing the susceptibilities of the two layers. 

The resonant nature of the response of each mode is expressed by the Lorentzian function:
\begin{equation}
\mathcal{L}^\textrm{mode}(\omega)=\chi_\textrm{max}^\textrm{mode} \frac{i \,  \omega_\textrm{mode} \,\Delta \omega_\textrm{mode} \,  }{(\omega_\textrm{mode}^2-\omega ^2)+ i \, \omega \, \Delta \omega_\textrm{mode} } ,
\label{Lorentzian} \end{equation} 
The partial responses $\mathcal{L}^\textrm{ac}$ and $\mathcal{L}^\textrm{op}$ peak at their resonance frequencies (Eq.~\ref{FMRfrequencies}). The $\mathcal{L}^\textrm{ac}$ and $\mathcal{L}^\textrm{op}$ are purely imaginary at their resonances and can be expressed as the dimensionless numbers:
\begin{equation}
\chi_\textrm{max}^\textrm{ac}=- i \frac{M_s} {H_J} \frac{\omega_\textrm{ac}}{\Delta \omega_\textrm{ac}} \textrm{~~and~~}
\chi_\textrm{max}^\textrm{op}=- i \frac{M_s} {H_J} \frac{\omega_\textrm{op}}{\Delta \omega_\textrm{op}} \label{chimax}.
\end{equation}
Several points are worth noticing in Eq.~\ref{ChiTotal}. \\
(i) As is well known, the optical mode responds to parallel pumping only, $H_x \parallel h^{rf}$ (the non-vanishing susceptibility terms are in the first column), while the acoustic mode responds only to transverse rf fields, $H_x \perp h^{rf}$ (second and third columns). \\
(ii) Near resonance, all susceptibility terms are essentially Lorentzian-like (Eq.~\ref{Lorentzian}). However the additional $\omega/\omega_\textrm{mode}$ factors in some of the non-diagonal susceptibility terms (Eq.~\ref{ChiTotal}) make them non-Lorentzian and let them vanish in the static limit. \\
(iii) The magnitudes of the layer-resolved susceptibilities of the two modes are very different. At low fields when $H_x \ll H_J \ll M_s$, we have $\chi^\textrm{ac}_\textrm{max} (H_x \rightarrow 0)\approx -\frac{i} {\alpha }\frac{H_x}{H_J} \sqrt{\frac{M_s}{H_J}}$ and 
$\chi^\textrm{op}_\textrm{max} (H_x \rightarrow 0) \approx -\frac{i} {\alpha } \sqrt{\frac{M_s}{H_J}}$. The susceptibility at low fields is much larger for the optical mode at resonance than for the acoustic mode at resonance. Conversely, the response of the optical mode progressively dies as one approaches saturation at $H_J$ while the resonant susceptibility of the acoustic mode attains it maximum with essentially $\chi^\textrm{ac}_\textrm{max} (H_x = H_J) \approx -\frac{i} {\alpha } \sqrt{\frac{M_s}{H_J}}$.

We highlight that since $\bar {\bar \chi}_\textrm{ac} h_\textrm{roll}^{rf} \neq \vec 0$,  
the lattice rotations are expected to excite the acoustic SW mode. It is the case when the strain is considered to be uniform within the SAF, which is a good approximation when it is much thinner than the SAW wavelength.
In contrast, $\bar {\bar \chi}_\textrm{op} h_\textrm{roll}^{rf}=\vec 0$, meaning that the lattice rotations do not excite the optical SW mode directly. This does not mean that the rolling fields could be disregarded for the optical mode, it would only be the case if  $\bar {\bar \chi}_\textrm{op} h_\textrm{tick}^{rf} \perp h_\textrm{roll}^{rf}$, which is in general not true. 

\section{Strain and rotation tensors of surface acoustic waves in Lithium niobate systems} \label{SAWsection}
The objective of this section is to describe the strain tensor and the lattice rotation tensor of Rayleigh SAWs travelling in LiNbO$_3$ systems. Targeting SAWs to operate at higher frequencies, one indeed replaces the traditional bulk piezoelectric substrate \cite{Kimura2018, Almirall2019} by a piezoelectric thin film, see Fig.~\ref{xyzXYZ}(a). In particular, thin films are grown on higher-acoustic-velocity substrates, typically forming an LiNbO$_3$/sapphire \cite{LaSpina2023} stack [Fig. \ref{xyzXYZ}(a)]. In such structure, the SAW is guided in the thin piezoelectric film, and the clamping effectively accelerates the acoustic waves while modifying their polarizations. As a bonus, the high confinement of the acoustic energy is expected to boost the SAW-SW interaction, with direct application interest. Therefore, it is of great interest to analyze the symmetry of this coupling in order to understand to what extent SWs can couple to the guided and non-guided SAWs. Here we consider the specific case of a SAF with the magneto-rotation and the magneto-elastic interactions being included on an equal footing.

We first study the case of an elastically isotropic non-piezoelectric medium mimicking LiNbO$_3$ ("iso" model). We then compare it to the Rayleigh wave propagating along X$_\textrm{LNO}$ crystalline direction of a semi-infinite LiNbO$_3$ substrate with a Z$_\textrm{LNO}$-oriented surface ("Z-cut" model). Finally, we will discuss the additional features that appear when the SAWs are guided in a thin film of Z-LNO grown on C-sapphire ("guided" model). 
\subsection{Rayleigh wave for an elastically isotropic substrate}\label{IsoLNOSection}
Following Ref. \cite{Thevenard2014}, the velocity $V$ of Rayleigh-type surface acoustic wave in an elastically isotropic medium obeys:
\begin{equation}
   \left(1-\frac{\rho  V^2}{C_{44}}\right)
   \left(1-\frac{C_{12}^2}{C_{11}^2}-\frac{\rho 
   {V}^2}{C_{11}}\right)^2-\frac{\rho ^2 V^4 \left(1-\frac{\rho 
   V^2}{C_{11}}\right)}{C_{11}^2} = 0 
   \label{RayleighEquation}
\end{equation}
where $\rho$ is the mass density, $C_{11}^\textrm{iso}$ the modulus of axial compression 
, $C_{44}^\textrm{iso}$ is the shear modulus 
, and $C_{12}^\textrm{iso}$ is the ratio of longitudinal stress $\sigma_{11}$ to transverse compression $\epsilon_{22}$, which reduces to $C_{12}^\textrm{iso}=C_{11}^\textrm{iso}- 2C_{44}^\textrm{iso}$ in case of isotropic elastic properties. The $^\textrm{iso}$ superscripts were omitted in the previous equation. The wave velocity is the smallest root of Eq.~\ref{RayleighEquation}. \cite{Thevenard2014}

The strict application of Eq.~\ref{RayleighEquation} with the material parameters of bulk \cite{Kovacs1990} LiNbO$_3$ would yield a SAW velocity of $3.37$ km/s. This poorly accounts for that of a Z-cut orientation for which the reported experimental values are 3.798-3.903 km/s 
\cite{Ciplys1999, Kushibiki2002}. 
By definition a Rayleigh wave has finite strains only in the X and Z directions for an X-propagating wave on a Z-cut substrate. The strains $\epsilon_{XX}$ and $\epsilon_{ZZ}$ have comparable amplitudes so that the best approximation for the modulus of axial compression is to take $C_{11}^\textrm{iso} = \left(C_{11}^\textrm{LNO} + C_{33}^\textrm{LNO}\right)/2$. Besides, since the active shear component of a Rayleigh wave is $\epsilon_{XZ}$, the natural choice for the shear modulus is $C_{44}^\textrm{iso} = C_{66}^\textrm{LNO}$, and $C_{12}^\textrm{iso} = C_{13}^\textrm{LNO}$. Making this substitution and applying Eq.~\ref{RayleighEquation}, one gets a Rayleigh mode velocity of 3.629 km/s, now in better agreement with experimental reports for the case of Z-cut LNO.
\begin{figure} 
\hspace{-2mm}
\includegraphics[width=8.5cm]{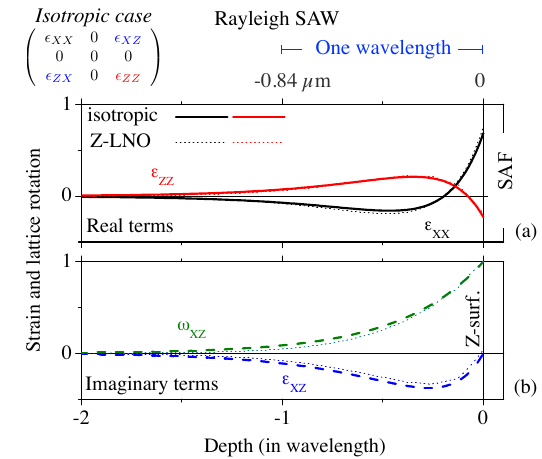}
\caption{Depth profile of the main strain components and lattice rotation component for a Rayleigh surface acoustic wave traveling on semi-infinite substrate, elastically isotropic (plain and dashed lines) or along the X crystalline direction of a Z-LNO substrate (dotted lines). (a) Terms that are real used as phase reference. (b) Terms that are imaginary (phase in quadrature). The inset shows the strain tensor in the elastically isotropic case. Additional non-vanishing terms arise for the Z-LNO case, as plotted in Fig.~\ref{specificZlno}. }
\label{IsotropicZLNO} 
\end{figure}
The thickness profile of the Rayleigh wave is defined by dimensionless penetration coefficients $q_1$ and $q_2$ \cite{Thevenard2014}:
\begin{equation}
    q_1=-\sqrt{1-\frac{\rho  V_r^2}{C_{11}}} ~\textrm{and}~
    q_2=-\sqrt{1-\frac{\rho  V_r^2}{C_{44}}}.
\end{equation} 
The thickness profile of the lattice displacement $\vec {\tilde {u}}(X, Z, t)$ is the sum of two partial waves:
\begin{equation}
    \left(
\begin{array}{c}
\tilde{u}_X= e^{-k q_1 z}-\frac{2 q_1 q_2 }{q_2^2+1}e^{-k
   q_2 z} \\
\tilde{u}_Y= 0 \\
\tilde{u}_Z= i q_1 \left(\frac{2 }{q_2^2+1}  e^{-k q_2
   z}- e^{-k q_1 z}\right) \\
\end{array}
\right) e^{i (\text{$\omega $t}-k X)} \label{IsoDisplacements}
\end{equation}
The lattice rotation tensor $\bar {\bar \omega}=\frac{1}{2} \left(\bar {\bar \nabla} \tilde{u}-\bar {\bar \nabla} \tilde{u}^T \right)$ and the strain tensor  $ \bar {\bar \epsilon}=\frac{1}{2} \left(\bar {\bar \nabla} \tilde{u}+\bar {\bar \nabla} \tilde{u}^T \right)$ can be straightforwardly obtained as the anti-symmetric part and the symmetric part of the displacement gradient \footnote{$\bar {\bar \omega}$ and $\bar {\bar \epsilon}$ were multiplied by $-i$ to set $\epsilon_{xx}$ as the phase reference. }. 

The strain and the lattice rotation tensors are plotted in Fig.~\ref{IsotropicZLNO} for material parameters meant to best mimic an isotropic equivalent of the Z-cut LiNbO$_3$ substrate. The material parameters of bulk~\cite{Kovacs1990} LiNbO$_3$ transform in $C_{11}^\textrm{iso}=213$ GPa, $C_{44}^\textrm{iso}=72$ GPa and $C_{12}^\textrm{iso}=65$ GPa, with $\rho=4628~\textrm{kg/m}^3$.
Since the displacements are strictly restricted to the $XZ$ plane and since for an X-propagating wave we have $\frac{\partial}{\partial y}=0$, all terms involving $Y$ cancel within $\bar{\bar \epsilon}$ and $\bar{\bar \omega}$. 

The non-vanishing terms and their phase relations are $\epsilon_{XX} \in \mathbb{R}$, $\epsilon_{ZZ} \in \mathbb{R}$, $\epsilon_{XZ} \in i \mathbb{R}^-$ and $\omega_{XZ} \in i \mathbb{R}^+$. The shear strain $\epsilon_{XZ}$ exactly vanishes at the free surface. Note also that the lattice rotation $\omega_{XZ}$ at the surface is greater than all the strain components, with foreseen implications for the amplitude of the rolling fields (Eq.~\ref{RollingFields}) and the tickle fields (Eq.~\ref{TickleFields}).

\subsection{Rayleigh wave for X-propagation and Z-cut LNO} \label{XZLNO}
It is interesting to analyze to what extent the above-studied Rayleigh SAW of an hypothetically isotropic substrate differs from that of the real (anisotropic and piezoelectric) material. The latter can be numerically calculated using the method described in ref.~\cite{LaSpina2023}, derived from the model of ref.~\cite{Laude2018}. This method solves for the resonant lattice displacements and the electrical potential across the depth of a multilayer, taking into account the elastodynamic equation, the Poisson equation and the constitutive relations of piezoelectricity (see \cite{LaSpina2023}), all using the material properties \cite{Kovacs1990} of bulk LNO.

\begin{figure} 
\hspace{-2mm} 
\includegraphics[width=8.cm]{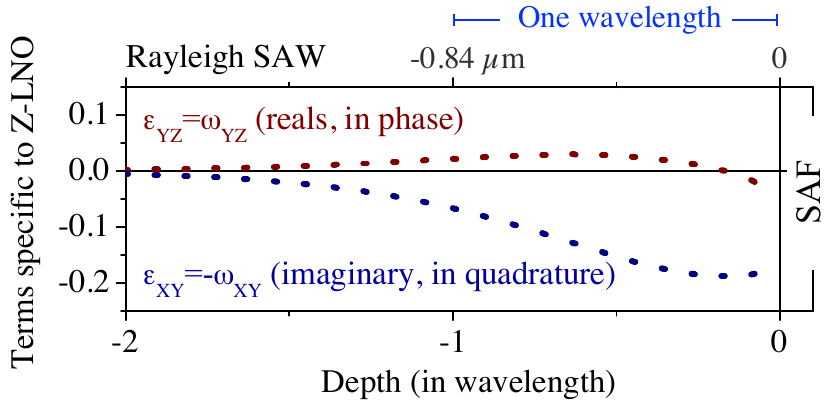}
\caption{Depth profile of the strain components and lattice rotation components that become non-vanishing when passing from an elastically isotropic substrate to a Z-LNO substrate for a SAW traveling along the X crystalline direction.}
\label{specificZlno} 
\end{figure}

For X-propagating SAWs on Z-LNO, the lowest frequency mode is calculated to have a velocity of 3.787 km/s, in agreement with experimental findings \cite{Ciplys1999, Kushibiki2002}. A finite displacement $u_Y$ appears as a consequence of the anisotropy of the rigidity tensor of LNO. However the wave polarization still lies essentially in the $XZ$ plane, and therefore it will be referred to as a Rayleigh mode. The depth profiles of its strain tensor and its lattice rotation tensor are displayed in Figs.~\ref{IsotropicZLNO} (pure Rayleigh terms) and \ref{specificZlno} (other terms). A comparison with the elastically isotropic case (also plotted in Fig.~\ref{IsotropicZLNO}) indicates that the large elastic deformations are still the pure Rayleigh terms $\epsilon_{XX}$, $\epsilon_{ZZ}$ and $\epsilon_{XZ}$, and that they are still smaller than the lattice rotation main term $\omega_{XZ}$. Actually, if looking at these sole four terms (Fig.~\ref{IsotropicZLNO}), it would be difficult to find major differences between the Rayleigh waves in real Z-LNO and in its elastically isotropic imitation. 
In the real materials $\epsilon_{YZ}$, $\epsilon_{XY}$, $\omega_{XY}$ and $\omega_{YZ}$ become finite although still remaining small \footnote{A minor difference is that thanks to piezoelectricity, the shear strain $\epsilon_{XZ}$ does no longer strictly vanish at the free surface.}  compared to that of the canonical Rayleigh wave [see the vertical scale in Fig.~\ref{specificZlno}]. We will discuss further below that these non-vanishing characters actually change the angular dependence of the rolling fields (Eq.~\ref{RollingFields}) and of the tickle fields (Eq.~\ref{TickleFields}), with consequences for the energy transfer between the SAW and the SWs.

\subsection{Rayleigh SAW guided in Z-LNO/sapphire system} \label{guidedSection}
\begin{figure} 
\hspace{-5mm}
\includegraphics[width=9.cm]{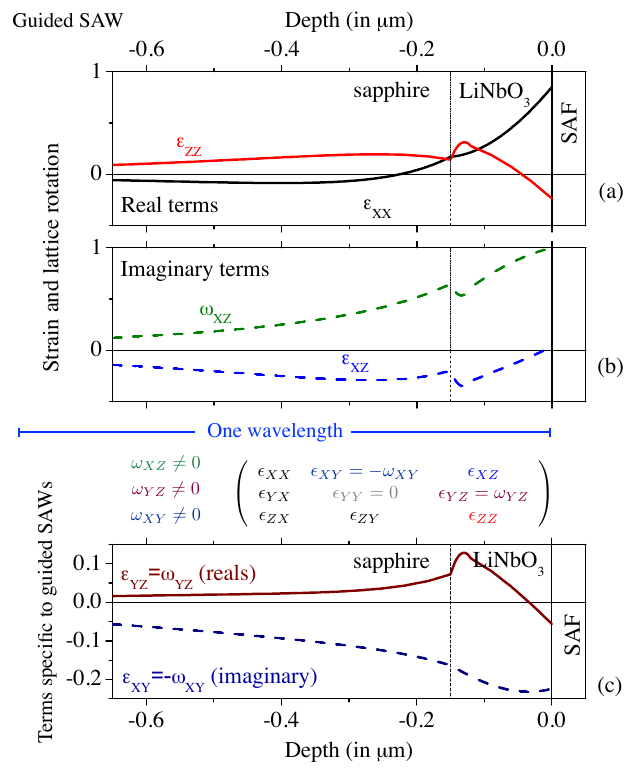}
\caption{Depth profile of the non-vanishing strain components and lattice rotation components for the lowest order surface acoustic wave traveling in a Z-LNO film (150 nm) on a C-sapphire substrate for $k=7.5 ~\textrm{rad}/\mu\textrm{m}$ (equivalently 5.88 GHz). (a-b) Terms that would already exist in an elastically isotropic, non-piezoelectric semi-infinite system like that of Fig.~\ref{IsotropicZLNO}. (c) General form of the strain and lattice rotation tensors, and depth profile of the components that are specific to this guided SAW. }
\label{ZLNO150nm} 
\end{figure}
Additional features emerge when considering guided piezoelectric thin films. 
We consider a 150-nm thick Z-LiNbO$_3$ film grown on a C-sapphire (Al$_2$O$_3$) substrate. The dispersion and elastic tensors of the Rayleigh SAW can again be numerically calculated using the model of \cite{Laude2018,LaSpina2023} and the elastic properties of sapphire \cite{Wachtman1960}. For a SAW frequency of 5.88 GHz, with corresponding wavelength of 833 nm, the guidance within the 150-nm thick LNO increases the wave phase velocity to 4.899 km/s. In addition to the acceleration effect, the guidance distorts the depth profiles (Fig.~\ref{ZLNO150nm}) of the elastic deformations and steers it towards net localization within the LiNbO$_3$ film. In addition, the mismatch of acoustic impedance between Al$_2$O$_3$ and LiNbO$_3$ induces some waviness within the depth profile of the elastic deformations at their interface (at $z=0$ in Fig.~\ref{ZLNO150nm}), in a manner recalling Sezawa waves. With respect to the impact on the coupling to SAWs, by reducing the thickness of the piezoelectric material, the confinement of the acoustic energy near the interface should lead to a more efficient magneto-mechanical coupling.

Just as in the pure Z-LNO case (§\ref{XZLNO}), there are non-zero strain/rotation coefficients related to the finite displacements in the Y direction. These terms can originate from the anisotropic character of Z-LNO. However a comparison between Fig.~\ref{specificZlno} and \ref{ZLNO150nm}(c) shows that shear strains and lattice rotations involving the transverse direction (i.e. $Y$) are slightly 
increased by the guidance of the wave, with a strongly enhanced $\epsilon_{YZ}$ recalling Love waves. The non-vanishing character of $\epsilon_{YZ}$ is known to result in different symmetry properties of magneto-elastic torques \cite{Kuss2021}. This aspect will be further discussed in section \ref{SAWSW-otherSubstrates}.      



\subsection{Comparison of elastic tensors at the surface}
For magnetic films much thinner than the wavelength of the SAWs, the magneto-elastic tickle fields and magneto-rotation fields can be anticipated from the surface strain and surface rotation. The magnetic film can also be assumed not to exert any back-action on the SAW. These surface deformations are compared in Table~\ref{TableI} for the "iso", "Z-cut" and "guided" models. It is noticeable that the piezo-electricity, the elastic anisotropy and the guidance do not alter significantly the deformations at the surface for the pure Rayleigh terms. The signs, amplitude and phase relations between $\epsilon_{XX}$, $\epsilon_{ZZ}$, $\epsilon_{XZ}$ and $\omega_{XZ}$ are very similar for the 3 material systems.
This similarity leads to conclude that Eq.~\ref{IsoDisplacements} is a satisfactory description of $\epsilon_{XX}$, $\epsilon_{ZZ}$ and $\omega_{XZ}$ of the acoustic waves of Rayleigh-type for the surface of our 3 material systems. While $\epsilon_{XZ}$ stays negligible in all configurations at the surface, the additional terms $\epsilon_{XY}$, $\epsilon_{YZ}$ and $\omega_{YZ}$ should be considered for bulk Z-LNO and, even more, for guided configurations \footnote{The lattice rotation $\omega_{XY}$ is also finite at the surface, but will not lead to a torque for the isotropic magnetic materials shall we shall consider hereafter.}.

\begin{table}[] 
\caption{Elastic strain tensors and lattice rotation tensors ($\{XYZ\}$ coordinate system) at the surface of the elastically isotropic LiNbO$_3$ ("iso" model §\ref{IsoLNOSection}), of X-prop-Z-cut LNO ("Z-cut" model §\ref{XZLNO}) and, of X-prop-Z-LNO(150nm)/C-sapphire "guided" model (§\ref{guidedSection}), for a Rayleigh surface acoustic wave.} 
 \begin{tabular}{|c|c|c|}
 \hline
 & Surface strain & Surface rotation  \\ 
 & $\bar {\bar \epsilon}$ & $\bar {\bar \omega}$  \\ \hline
iso & $\left(
\begin{array}{ccc}
 0.69 & 0 & 0 i \\
 0 & 0 & 0 \\
-0 i  & 0 & -0.23\\
\end{array}
\right)$  & $\left(
\begin{array}{ccc}
 0 & 0 & i \\
 0 & 0 & 0 \\
-i  & 0 & 0\\
\end{array}\right)$          \\ \hline
Z-cut & $\left(
\begin{array}{ccc}
 0.76 & -0.17 i & 0.04 i \\
 -0.17 i & 0 & -0.05 \\
 0.04 i & -0.05 & -0.18 \\
\end{array}
\right)$  & $\left(
\begin{array}{ccc}
 0 & 0.17 i & i \\
 -0.17 i & 0 & -0.05 \\
 -i & 0.05 & 0 \\
\end{array}
\right)$          \\ \hline
guided & $\left(
\begin{array}{ccc}
 0.85 & -0.22 i & 0.04 i \\
 -0.22 i & 0 & -0.06 \\
 0.04 i & -0.06 & -0.24 \\
\end{array}
\right)$  & $\left(
\begin{array}{ccc}
 0 & 0.22 i & i \\
 -0.22 i & 0 & -0.06 \\
 -i & 0.06 & 0 \\
\end{array}
\right)$          \\ \hline
 \end{tabular} \label{TableI}
 \end{table}

\subsection{Amplitudes of the deformations}
So far we described the polarization of the acoustical waves (Table \ref{TableI}). It is also worth discussing the amplitudes of the elastic deformations that can be effectively achieved. In practice the SAW is almost always generated/collected by interdigital transducers (IDTs) forming a delay line.
Assuming that the IDTs are connected to an rf source by lossless cables, at most half of the electrical power $P_\textrm{rf}$ can be transmitted to elastodynamic power $P_\textrm{SAW}$ in the form of a SAW travelling in the propagation direction imposed by the delay line. Neglecting resistive losses, the microwave acoustic power below the input IDT reads:     
\begin{equation}
P_\textrm{SAW}=\frac{1}{2} \, P_\textrm{rf} \, (1-||S_{11}||^2) ,\end{equation}  
where $S_{11}$ is the reflection coefficient of the input IDT.

For a Rayleigh wave propagating in an isotropic medium, if the acoustical power $P_\textrm{ac}$ is spread over a wavefront of acoustic aperture $w_\textrm{ap}$ (for width), the so-called total displacement \cite{royer1999elastic} is \begin{equation} 
||u_N|| = \sqrt{2 P_\textrm{SAW} / (w_\textrm{ap} \rho \omega V_r^2)} \label{Un}.
\end{equation}  
The typical strain is $k\,.||u_N||$ where $k=\omega / V_r$ is the SAW wavevector. 

Experiments typically use an electrical power of $P_\textrm{rf}=1$ mW, IDTs of reflection coefficients $||S_{11}|| = 0.95$ at a frequency of 600 MHz and an acoustical aperture of $w_\textrm{ap}=35~\mu$m. 
These numbers lead to a typical strain of $2.3 \times 10^{-5}$. 
Better electrical matching can lead to a reduced $||S_{11}||$ and therefore much larger elastic deformation. From now on, we arbitrarily choose $\omega_{xz}=10^{-5} i$ and scale the strain components accordingly. 

\section{Surface acoustic waves dissipation to the SAF} \label{sectionCoupling}

\subsection{Power supplied to the magnetic material} 
 
The lattice deformations create space-time-varying effective fields $\mathcal R e \left((h^{rf}_\textrm{tick}+h^{rf}_\textrm{roll}) e^{i (\omega t - k_{SAW} X)}\right)$ that in turn excite magnetization precession. The SAW power absorbed by the magnetic damping is usually calculated using the effective field approach \cite{Dreher2012}. The time-averaged power transmitted to the magnetic layers at position $X$ (per unit surface where the SAW is present) is:
\begin{equation}
    \Delta P(X) =-\frac{\mu_0 \omega  }{2} \textrm{Im} \left(\sum_{layers} t_\ell (h^{rf})^\dagger \cdot\bar {\bar {\chi}}_\ell \cdot h^{rf} \right) \textrm{~~~in~W/m}^2
    \label{DeltaP}
\end{equation} 
where the dagger means transpose-conjugate, and $h^{rf}=(h_\textrm{tick}^{rf} +  h_\textrm{roll}^{rf})$ is the (complex-valued) total effective rf field. 
Being a power, $\Delta P(X)$ is a positive scalar that is independent from the chosen reference frame. Under the assumption of Eq.~\ref{WorkingAssumption} the time-averaged power transmitted to
the magnetic layers is spatially uniform: $\Delta P(X)=\Delta P$ \footnote{Note that if removing the assumption of Eq.~\ref{WorkingAssumption} the expression of $\Delta P(X)$ would still hold provided that the magnetic susceptibility $\bar {\bar \chi}$ is expressed for the wavevector $k_\textrm{SAW}$}.  We shall calculate it in the reference frame $xyz$ of the magnetic field. For this, we need to express the tickle field and the rolling field associated to each of the SAWs of our 3 material systems (Table \ref{TableI}) in the reference frame of the magnetic field. This is done by using Eq.~\ref{TickleFields} and \ref{RollingFields} and expressing the elastic tensors as:
$\bar {\bar \epsilon}_{xyz}= \bar {\bar R} \cdot \bar {\bar \epsilon}_{XYZ} \cdot \bar {\bar R}^T$ and $\bar {\bar \omega}_{xyz}= \bar {\bar R} \cdot \bar {\bar \omega}_{XYZ} \cdot \bar {\bar R}^T$ where $^T$ is the transposition operation.

\subsection{Magnitudes of the different channels of SAW to SW energy transfer}
The time-averaged power transmitted to the magnetic layers $\Delta P$ is a function of the experimental conditions $H_x$, $\omega$, $\varphi$ and of the magnetic material properties $\gamma_0$, $M_s$, $H_J$, $B_{1,2}$ and $\alpha$. The two SAF SW eigenmodes (acoustic and optical) are independent, such that the power supplied to the magnetic material (Eq.~\ref{DeltaP}) can be expressed as the sum of the contribution of each SW mode as $\Delta P=\Delta P_\textrm{ac}+\Delta P_\textrm{opt}$. The contribution of each SW mode scales with the corresponding susceptibilities, which strongly depend on the applied field $H_x/H_J$ through Eq.~\ref{chimax} as was discussed in point (iii) of section \ref{SusceptibilitySection}.

\begin{figure} 
\hspace{-5mm}
\includegraphics[width=9.cm]{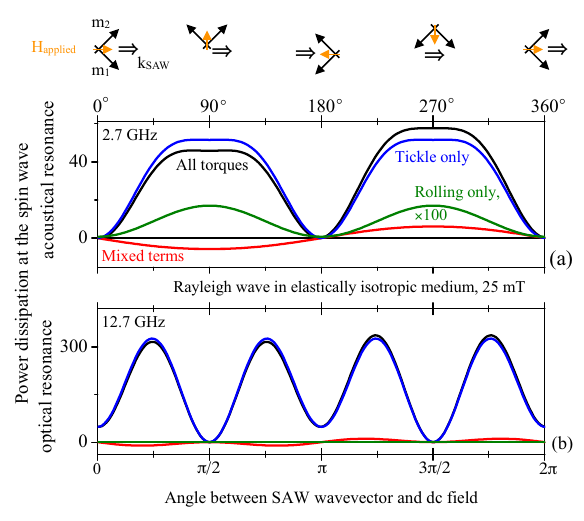}
\caption{Contributions of the pure magneto-elastic term $(h_\textrm{tick}^{rf} )^\dagger \cdot \bar {\bar {\chi}}  \cdot (h_\textrm{tick}^{rf})$ (blue), the pure magneto-rotation term $(h_\textrm{roll}^{rf} )^\dagger \cdot\bar {\bar {\chi}} \cdot(h_\textrm{roll}^{rf} )$ (green, magnified by a factor of 100) and the mixed term $(h_\textrm{roll}^{rf} )^\dagger \cdot\bar {\bar {\chi}} \cdot(h_\textrm{tick}^{rf} )$ + $(h_\textrm{tick}^{rf} )^\dagger \cdot\bar {\bar {\chi}} \cdot(h_\textrm{roll}^{rf} )$ (red) to the total power dissipation (black). A Rayleigh wave is assumed in an elastically isotropic, non-piezoelectric semi-infinite system like that of Fig.~\ref{IsotropicZLNO}. The applied dc field is 25 mT. Panel (a) and (b) are respectively for applied frequencies that are resonant with the SW acoustic or optical modes.}
\label{25mTMagnitudesOfTheDifferentTerms}
\end{figure} 
In contrast, the contributions of the tickle field and of the rolling field are coupled. Indeed, in addition to the pure magneto-elastic term $(h_\textrm{tick}^{rf} )^\dagger \cdot\bar {\bar {\chi}} \cdot(h_\textrm{tick}^{rf})$ and the pure magneto-rotation term $(h_\textrm{roll}^{rf} )^\dagger \cdot\bar {\bar {\chi}} \cdot(h_\textrm{roll}^{rf} )$ that are unconditionally positive, one should also consider the "mixed" terms $(h_\textrm{roll}^{rf} )^\dagger \cdot\bar {\bar {\chi}} \cdot(h_\textrm{tick}^{rf} )$ and $(h_\textrm{tick}^{rf} )^\dagger \cdot\bar {\bar {\chi}} \cdot(h_\textrm{roll}^{rf} )$ that can have arbitrary signs. Examples of the expected contributions of these terms in the "iso" model are plotted in Fig.~\ref{25mTMagnitudesOfTheDifferentTerms} and ~\ref{100mTMagnitudesOfTheDifferentTerms} for applied fields $H_x = 0.2 H_J$ and $H_x = 0.8 H_J$. 

\subsection{Reciprocity of the different channels of SAW to SW energy transfer} 
\begin{figure} 
\hspace{-5mm}
\includegraphics[width=9.cm]{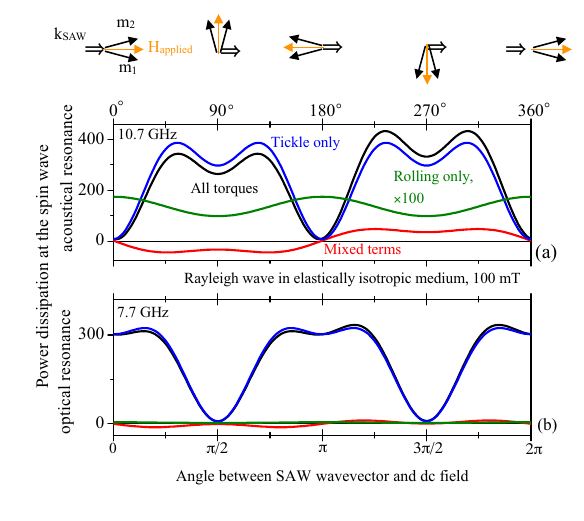}
\caption{Contributions of the pure magneto-elastic term $(h_\textrm{tick}^{rf} )^\dagger \cdot \bar {\bar {\chi}}  \cdot (h_\textrm{tick}^{rf})$ (blue), the pure magneto-rotation term $(h_\textrm{roll}^{rf} )^\dagger \cdot\bar {\bar {\chi}} \cdot(h_\textrm{roll}^{rf} )$ (green, magnified by a factor of 100) and the mixed term $(h_\textrm{roll}^{rf} )^\dagger \cdot\bar {\bar {\chi}} \cdot(h_\textrm{tick}^{rf} )$ + $(h_\textrm{tick}^{rf} )^\dagger \cdot\bar {\bar {\chi}} \cdot(h_\textrm{roll}^{rf} )$ (red) to the total power dissipation (black). A Rayleigh wave is assumed in an elastically isotropic, non-piezoelectric semi-infinite system like that of Fig.~\ref{IsotropicZLNO}. The applied dc field is 100 mT. Panel (a) and (b) are respectively for applied frequencies that are resonant with the SW acoustic or optical modes.} 
\label{100mTMagnitudesOfTheDifferentTerms}
\end{figure}

It is also interesting to identify which terms lead to reciprocal or non-reciprocal transduction, as illustrated for the "iso" model in Fig.~\ref{25mTMagnitudesOfTheDifferentTerms} and ~\ref{100mTMagnitudesOfTheDifferentTerms}. Thanks to the assumption of Eq.~\ref{WorkingAssumption}, the non-reciprocity discussed here only originates from an helicity mismatch. The properties of the rotation operator ensure that: 
\begin{equation}
\left\{
\begin{array}{ll}
\epsilon_{xx,yy,zz, xy}(\varphi) & = ~~\epsilon_{xx,yy, zz, xy} (\varphi + \pi) \\
\epsilon_{xz, yz}(\varphi) &= - \epsilon_{xz, yz} (\varphi + \pi)  \\
\omega_{xz, yz}(\varphi) &= - \omega_{xz, yz} (\varphi + \pi) \\
\end{array}
\right.
\label{NRrules}
\end{equation}
In the "iso" model, we have in addition $\epsilon_{xz}=0$ at the surface. Besides, in in-plane magnetized isotropic SAF $\epsilon_{zz}$ and $\omega_{xy}$ do not provide any magnetic torque.

As long as the magnetic layer is much thinner than the SAW peneration depth, it will be reasonable to neglect the $\epsilon_{xz}$ torque. In the "iso" model the tickle field (Eq.~\ref{TickleFields}) lies in the sample plane (no $z$ component) and obeys $h^{rf}_\textrm{tick}(\varphi+\pi)=h^{rf}_\textrm{tick}(\varphi)$, while the rolling field (Eq.~\ref{RollingFields}) is along $z$ and obeys $h^{rf}_\textrm{roll}(\varphi+\pi)=- h^{rf}_\textrm{roll}(\varphi)$. As a result, the 
pure magneto-elastic term $(h_\textrm{tick}^{rf} )^\dagger \cdot \bar {\bar {\chi}} \cdot (h_\textrm{tick}^{rf})$ and the pure magneto-rotation term $(h_\textrm{roll}^{rf} )^\dagger \cdot\bar {\bar {\chi}} \cdot (h_\textrm{roll}^{rf} )$ are both reciprocal [if taken independently they would yield $\Delta P (\varphi)= \Delta P (\varphi + \pi)$]. 

When the material is both subject to magneto-elastic interactions and magneto-rotation interactions, the "mixed" terms have to be taken into account. Indeed the interplay between the tickle field and the rolling field leads to helicity mismatches that differ for $\varphi$ and $\varphi+\pi$. This induces a non-reciprocity of the excitation of spin waves,  we have $\Delta P (\varphi) \neq \Delta P (\varphi + \pi)$ even in the simplest "iso" case. This mechanism is analogous to that invoked in single layer ferromagnets: the constructive/destructive interplay between on the one hand the magneto-elastic torques arising from the longitudinal strain, and on the other hand, the shear strain components $\epsilon_{XZ}$ \cite{Hernandez2020} and the magneto-rotation torque related to the lattice rotation tensor $\omega_{XZ}$ \cite{Xu2020,Kuess2020}.

The additional strain components present in the realistic elastic models will understandably complicate the analysis of non-reciprocity.
Let us thus first discuss the dependence of the SAW to SW coupling on the angle between the SAW wavevector and the applied magnetic field, first in the "iso" case.
Note that in these examples, but also for the "Z-cut" and "guided" models (not shown), the response due to the tickle field is by far the largest contribution. This is related to the fact that we have chosen a strongly magnetostrictive material, therefore $\frac{2B_1} {M_s} \gg  \mu_0 M_s$ while the elastic rotations $\omega_{xz}$ and the elastic strain components ($\epsilon_{xx}$, etc.) are of the same order. 

\section{SAW-SW coupling in the elastically isotropic case} \label{SAWSW-iso} 
\begin{figure*} 
\hspace{-5mm}
\includegraphics[width=16.cm]{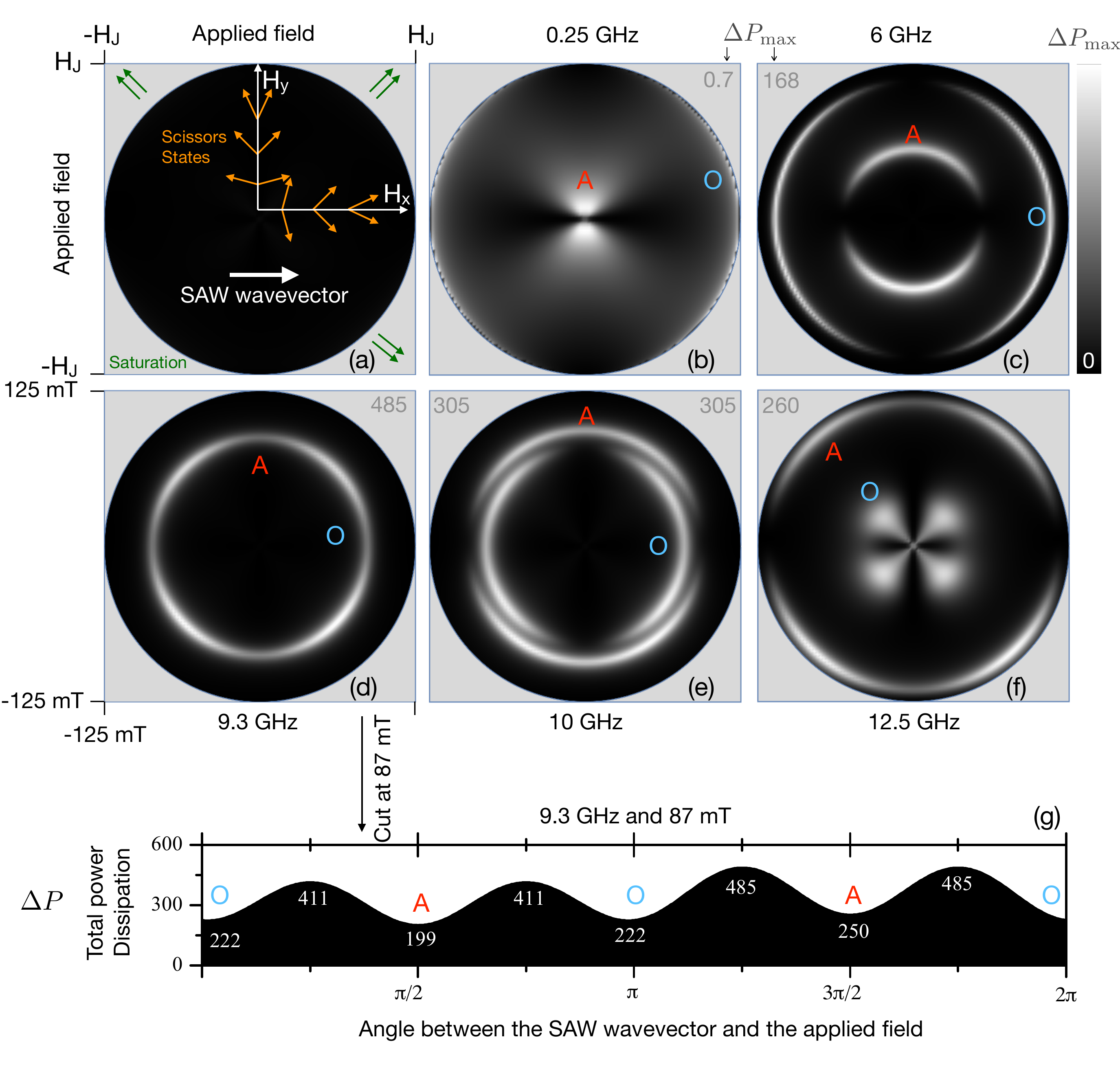}
\caption{Dependence within the plane of the applied field of the power absorbed (Eq.~\ref{DeltaP}) by a CoFeB-based SAF in the scissors state. The energy is supplied by the elastic deformations and lattice rotations at the surface of an elastically isotropic equivalent of Z-LiNbO$_3$ carrying a Rayleigh surface acoustic wave (Eq.~\ref{IsoDisplacements}, "iso" model). The magnetic parameters are $\mu_0 H_J=125$ mT 
, $\mu_0 M_s= 1.7$ T
, $\alpha=0.01$, $B_1=B_2=-7.6$ MJ/m$^3$. The "A" and "O" symbol label the contributions of the acoustic and optical SWs. (a) Schematics of the static magnetizations of the SAF. Power dependances (b) for a low frequency SAW, (c) for a SAW at 6 GHz, (d) for SAW at 9.3 GHz, (e) for a SAW at 10 GHz and (f) for a SAW at 12.5 GHz. The numbers in the upper corners of each panel indicate the maximum absorption. (g) Angular dependence of $\Delta P$ when $\omega_\textrm{ac}=\omega_\textrm{op}$  that highlights the omnidirectional coupling in a SAF.}
\label{GreyImagesISO} 
\end{figure*}

Fig.~\ref{25mTMagnitudesOfTheDifferentTerms} and \ref{100mTMagnitudesOfTheDifferentTerms} evidenced that the $\varphi$-dependence of $\Delta P$ at the acoustic and optical SW resonances depends strikingly on the applied field strength. It can feature 2-fold or 4-fold behaviors, with in addition a variable non-reciprocity amplitude, as shown in Fig.~\ref{GreyImagesISO} for a few representative frequencies and for all values of the applied field leading to non-collinear ("scissors") states. The behavior can be cast in three regimes.

\subsection{Low applied frequencies: weak coupling}
At low SAW frequency [Fig.~\ref{GreyImagesISO}(b)], $\Delta P$ is bound to be very low because of the $\omega$ prefactor in Eq.~\ref{DeltaP}. It is therefore a situation of little practical interest. The resonant condition for the SW acoustic (or optical) mode requires a low (or high) applied field, and the SAF is in a quasi-antiparallel (q-AP) (or quasi-parallel [q-P]) state with magnetizations almost perpendicular (or parallel) to the applied field.  
The angular dependence of $\Delta P_\textrm{op}(\varphi)$ recalls that of a single layer \cite{Weiler2011}. The coupling vanishes for $\varphi = n \pi /2,~n \in \mathbb Z$ and is maximal for $\varphi = \pi /4 + n \pi /2,~n \in \mathbb Z$. In contrast, $\Delta P_\textrm{ac}(\varphi)$ vanishes only for $\varphi = n \pi, ~n \in \mathbb Z$, though it remains also very weak.

\subsection{Intermediate frequencies with resonances in scissors configurations: inductive-like behavior}
Increasing the frequency yields a much higher coupling [Fig.~\ref{GreyImagesISO}(c-e)], and the resonances are met for "archetypal" scissors states (i.e. very distinct from the collinear states). A coupling to the SW acoustic mode happens whenever $\varphi \neq n\pi,~ n \in \mathbb Z $. A coupling to the SW optical mode occurs whenever $\varphi \neq \pi/2 + n\pi, ~n \in \mathbb Z$. In this frequency range, the magneto-elastic interaction acts very much as if the SAF excitation was done using a linearly polarized induction field parallel to the SAW wavevector. Indeed the parallel (perpendicular) pumping of a SAF using an rf magnetic field excites only the optical (acoustic) SW mode \cite{Zhang1994}. 

Interestingly, there exists a peculiar combination of field (87 mT) and frequency (9.3 GHz) for which the SW modes are both resonant [Figs.~\ref{saf}(a) and~\ref{GreyImagesISO}(d)], leading to no extinction of the SAW-SW coupling. 
Fig.~\ref{GreyImagesISO}(g) shows a cut at 87 mT that illustrates this behavior. At this specific applied field, the efficiencies of the excitation of the two eigenmodes have complementary angular variations, which prohibits absorption vanishing. In addition, this $\Delta P$ angular dependence is independent of the symmetry of the  tickle field associated to the magneto-elastic interaction, and of the rolling field associated with the magneto-rotation interaction.

\subsection{High frequencies with resonances for quasi-collinear states: single-layer-like behavior}
Finally at very high SAW frequency [Fig.~\ref{GreyImagesISO}(f)], the resonant condition for the acoustic mode requires a high applied field, so that the SAF is in a quasi parallel state. As a result, the acoustic mode now responds in a manner recalling that of a single layer, \textit{i.e.}, maximal for $\varphi = \pi /4 + n \pi /2,~n \in \mathbb Z$ and vanishing for $\varphi = 0 + n \pi /2$. Note that some amplitude non-reciprocity at the acoustic SW frequency is still visible as resulting from the mixed terms. See, for example, Fig.~\ref{GreyImagesISO}(f) where the power absorption is weaker in the $[0, \pi]$ interval than in the $[\pi, 2 \pi]$, as in the single layer case.

Conversely, the resonance with the optical SW mode requires now a low field, hence a quasi antiparallel state. The maximum couplings are also obtained for $\varphi = \pi/4 + n \pi/2, ~n \in \mathbb Z$ and the efficiency map resembles the four-leaf clover pattern that is classically \cite{Dreher2012} encountered in single layers subject to a pure longitudinal strain ($\epsilon_{xx}$). 



\section{Comparison of the three substrate models} \label{SAWSW-otherSubstrates}

\begin{figure} 
\hspace{-5mm}
\includegraphics[width=9.cm]{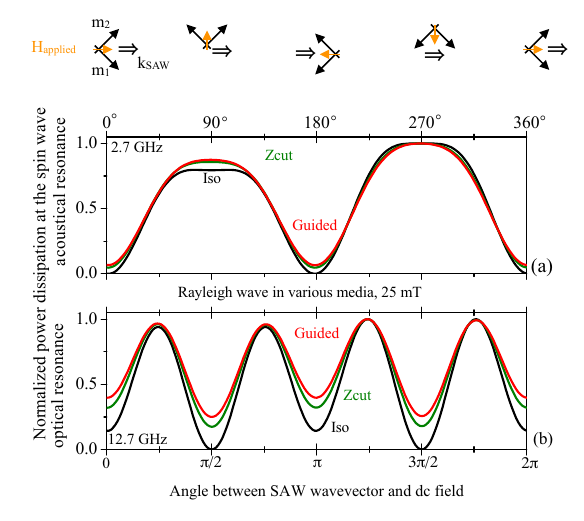}
\caption{Dependence of the power absorbed with the angle between the magnetic field and SAW wavevector for the three studied substrates (elastically isotropic equivalent of Z-LiNbO$_3$ ["iso" model], X-propagating Z-cut LiNbO$_3$ ["Z-cut" model] and guided LiNbO$_3$/sapphire ["guided" model]. The CoFeB-based SAF is in the scissors state. The applied dc field is 25 mT. Panel (a) and (b) are respectively for applied frequencies that are resonant with the SW acoustic or optical mode. The data have been normalized by the maximal absorption.
}
\label{ComparisonSubstrates25mT} 
\end{figure}

\begin{figure} 
\hspace{-5mm}
\includegraphics[width=9.cm]{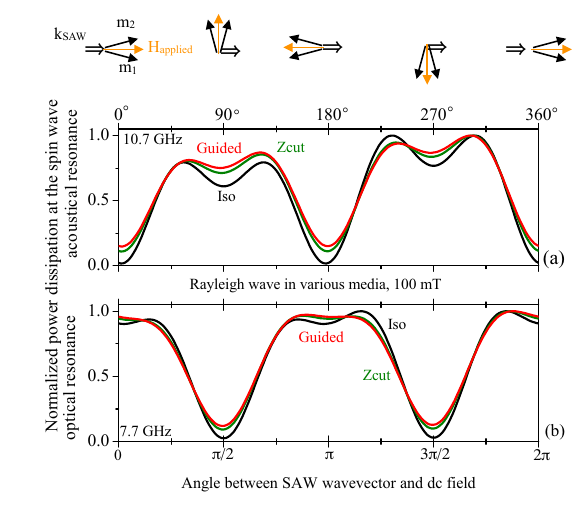}
\caption{Dependence of the power absorbed with the angle between the magnetic field and SAW wavevector for the three studied substrates (elastically isotropic equivalent of Z-LiNbO$_3$ ["iso" model], X-propagating Z-cut LiNbO$_3$ ["Z-cut" model] and guided LiNbO$_3$/sapphire ["guided" model]. The CoFeB-based SAF is in the scissors state. The applied dc field is 100 mT. Panel (a) and (b) are respectively for applied frequencies that are resonant with the SW acoustic or optical mode. The data have been normalized by the maximal absorption.
}
\label{ComparisonSubstrates100mT} 
\end{figure}

We now study to what extent the anisotropy of the elastic properties and the piezoelectric character of the substrate may change the angular dependence of power absorption by the spin waves, and also whether this is more prominent in the guided configuration. The angular dependence of the absorption within the 3 models of substrates are compared in Fig.~\ref{ComparisonSubstrates25mT} and \ref{ComparisonSubstrates100mT} for two different magnitudes of the applied field. The data were normalized to ease the comparison. 

In the case of the elastically isotropic substrate ("iso" model, black curves in Fig.~\ref{ComparisonSubstrates25mT} and \ref{ComparisonSubstrates100mT}), the absorption by the acoustic SW mode \textit{strictly} extinguishes at $\varphi=0$. That of the optical SW mode \textit{strictly} extinguishes at $\varphi=\pi/2$. As a result, the total absorptions at both resonances were always almost \footnote{The non-resonant absorption due to the other SW mode, although being very weak, does never vanish.} vanishing at some angles during angular scans of the applied field, except for the very specific applied field leading to a doubly resonant situation with $\omega_{ac}\approx \omega_{opt}$ like in Fig.~\ref{GreyImagesISO}(d).

In stark contrast, this extinction of the absorption at some angles no longer occurs when taking into account the elastic anisotropy and the piezolectricity of the substrate ("Z-cut" and "guided" models), indicated by the blue and red curves in Fig.~\ref{ComparisonSubstrates25mT} and \ref{ComparisonSubstrates100mT}. 
The resulting new interaction channels couple the magnetization eigenmodes to the elastic waves for all applied field directions. 
This suppression of the extinction can be understood from the single layer case where the field orientation dependence of the magneto-elastic coupling is well-known \cite{Dreher2012}. Indeed in single layer films the (here dominating) longitudinal strain $\epsilon_{xx}$ leads to a 4-fold coupling that extinguishes at $\varphi= 0 + \mathbb ~n \pi/2$,$~n \in \mathbb Z$. 
However the anisotropic character of the substrates results in a finite in-plane shear strain $\epsilon_{xy}$ (see Table \ref{TableI}) known to lead to a 4-fold  efficiency that is maximum at $n \pi/2$, precisely when the longitudinal strain looses efficiency, thereby suppressing any possibility of angular extinction, as discussed in Ref. \cite{Kuss2021}. Besides, the piezoelectricity present in the "Z-cut" and the "guided" models allows for finite out-of plane shear strains $\epsilon_{xz}$ and $\epsilon_{yz}$ at the free surface. In single layer films, these out-of-plane shear strains lead to 2-fold maximum efficiencies at $n\pi$ and $\pi/2 + n\pi$, respectively.

\section{Summary and conclusion}

We have analysed how surface acoustic waves travelling in a substrate transfer energy irreversibly to a SAF cap. Our study was conducted under several simplifying assumptions.
(i) We considered that the SAF cap does not perturb the elastic wave traveling in the underlying material. In other words, we disregarded any elastic back-action and any magneto-elastic back-action.
(ii) In addition, we have considered that the spin waves have a vanishing group velocity and that their susceptibility at finite wavevector does not differ from the uniform one.

We obtained a compact form for the layer-resolved susceptibility tensor. The expression separates the contributions of the SW acoustic eigenmode and of the SW optical eigenmode. The eigenmodes react to the tickle field associated to the magneto-elastic interaction, and to the rolling field associated with the magneto-rotation interaction.

 The model of a Rayleigh surface acoustic wave traveling in a elastically isotropic substrate is first recalled. Under these assumptions, the surface strain has only two non-vanishing components, the longitudinal strain and the (here magnetically inactive) out-of-plane strain. They are smaller but of the same order as the lattice rotations at the surface. We then analysed the Rayleigh surface acoustic wave traveling in a realistic material (X-propagating, Z-cut LiNbO$_3$) as well as in a similar material but in the form of a film clamped by a sapphire substrate (guided geometry). Additional strain components (in-plane and out-of-plane shear strains) emerge, allowing for additional SAW-SW coupling channels. Besides, we confirm that the clamping of the guided wave substantially accelerates the elastic wave, and changes the thickness profile of the elastic deformations though it marginally affects the largest strain components at the surface.

The SAW-SW coupling were evaluated by the energy supplied by the elastic waves to the magnetic degree of freedom. The magneto-elastic effects, especially the longitudinal compression term, dominate for our strongly magnetostrictive CoFeB-based SAF. However the rolling fields must still be considered to correctly account for the energy transfer and the non-reciprocity thereof. Indeed the interplay between the rolling field and the tickle field induces a substantial "mixed" term in the energy supplied by the SAW to the SW, and this term can be the main source of the amplitude non-reciprocity of the SAW-SW transduction. We also emphasize that the non-reciprocity originated from the SAF SW dispersion relation, as discussed in Ref. \cite{millo_unidirectionality_2023-1}, is not taken into account in the present work.  

The dependence of the energy transfer versus the angle between the applied magnetic field and the wavevector of the SAW was finally studied for an elastically isotropic substrate. It appears that at low and high applied fields, when the SAF is in close-to-collinear magnetic configurations, the angular dependence of the energy transfer can be understood by analogy with the Dreher-Weiler model \cite{Dreher2012} of a single magnetic layer subjected to a purely longitudinal strain. Conversely in scissors state, the behavior bears some similarity with the one that would be induced by an rf magnetic field oriented along the propagation direction of the SAW. 

An important outcome of this study is the symmetry analysis of the SAW-SW coupling. The symmetries of the two SW modes of SAF are complementary, giving rise to a large coupling \textit{whatever the field direction}. This is specially interesting for applications as it lifts geometry limitations. In addition, a slightly more complex behavior is obtained when allowing for elastic anisotropy and by taking into account the piezoelectricity of the substrate, in contrast to the isotropic case. It results in additional tensor components. The main feature is that the resulting symmetry of the magneto-elastic and magneto-rotation driving fields leads to new interaction channels allowing omnidirectional coupling. 

\begin{acknowledgments}

 This work was supported by public grants overseen by the French National Research Agency (ANR) as part of the “Investissements d'Avenir” and France 2030 programs (Labex NanoSaclay, reference: ANR-10-LABX-0035, project SPICY. Also: PEPR SPIN, references: ANR 22 EXSP 0008 and ANR 22 EXSP 0004). R. L. S, L. L. S, and F. M acknowledge the French National Research Agency (ANR) under Contract No. ANR-20-CE24-0025 (MAXSAW).

\end{acknowledgments}

\bibliography{references}

\end{document}